\documentclass[aps,prl,twocolumn,showpacs,preprintnumbers,amsmath,amssymb,groupedaddress,longbibliography]{revtex4-2}
\usepackage{hyperref}
\hypersetup{colorlinks=true, citecolor=blue, urlcolor=blue, linkcolor=blue,urlcolor=cyan}
\usepackage{color}
\usepackage{graphicx}
\usepackage{dcolumn}
\usepackage{braket}
\usepackage{amsthm}
\usepackage[caption=false]{subfig}

\def\be{\begin{equation}}
\def\ee{\end{equation}}
\def\ba{\begin{eqnarray}}
\def\ea{\end{eqnarray}}

 \def\ba{{\bar{\alpha}}}

\def\Tr{{\text{Tr}}}

\def\CH{{\mathcal{H}}}
\def\CM{{\mathcal{M}}}

\def\CB{{\mathcal{B}}}

\begin{document}

\title{Pure states for subregions in gravity and their entanglement entropy}

\author{Zixia Wei}
\affiliation{\it 
Society of Fellows, Harvard University, Cambridge, MA 02138, USA
}

\begin{abstract}
It is proposed that spatial subregions in quantum gravity can be assigned pure states, rather than mixed reduced density matrices. The state is prepared by a partially frozen gravitational path integral, in which a spacetime subregion containing the spatial subregion is fixed while the field configurations and ambient geometry are summed over. In the semiclassical regime, we further propose a holographic prescription for the entanglement entropy of bipartitions of this state, with a frozen-region analogue of the homology constraint. The prescription satisfies nontrivial self-consistency conditions, including strong subadditivity, complementarity, and entanglement wedge nesting, and reproduces several known entropy formulas in holography and gravity as special cases. The construction suggests an observer-dependent entanglement wedge labeled by the frozen subregion.

\end{abstract}

\maketitle

{\bf Introduction.---} 
In a closed quantum many-body system, the entire system is described by a pure state, while a proper subsystem is generically described by a mixed reduced density matrix, reflecting the lack of information about the degrees of freedom outside the subsystem. 
What is the corresponding statement for a subregion of a universe with gravity? At the semiclassical level, one expects little qualitative difference: tracing over exterior degrees of freedom yields a mixed state for the subregion. In the quantum gravity regime, however, the holographic principle \cite{tHooft93,Susskind94,Maldacena97} suggests that the information which appears to lie outside the subregion may instead be encoded on its boundary. This raises the possibility that the subregion itself admits a pure-state description. 

In this Letter, we make this intuition precise by proposing that each spatial subregion in quantum gravity can be associated with a pure state prepared by a gravitational path integral (GPI) with fixed subregion data. 
To characterize this pure state, we further propose a Ryu--Takayanagi-like prescription \cite{RT06,RT06b} for the entanglement entropy (EE) associated with its bipartitions, and verify that the prescription obeys nontrivial consistency conditions, including strong subadditivity (SSA), complementarity, and entanglement wedge nesting. For simplicity, we restrict to time-reflection-symmetric cases.

Let us begin with the question of what it means to specify a subregion in quantum gravity. Some data must clearly be supplied. A natural choice is the intrinsic geometry of the subregion, although other choices are possible. 
For instance, one may specify only geometry of its boundary. In any case, these specified data will be held fixed and enter the GPI as input boundary conditions for the GPI. 
As we will see, this may be viewed as subregion generalization gravitationally prepared states \cite{CGM20}. We now formulate this generalization and show that the resulting object is naturally interpreted as a pure state associated with the subregion.

~\par 
{\bf Gravitationally prepared states for subregions.--} 
A gravitationally prepared state on a closed spatial $\Sigma$ slice is proposed in \cite{CGM20} and defined as follows. Consider a $D$-dimensional ($D$-D) Einstein gravity coupled to a QFT described by the Euclidean action $I_E$, and start with a spacetime tube $\Sigma \times I$ on which the gravity is turned off. Let us use $\tau \in [-\epsilon,+\epsilon]$ to parameterize the direction along $I$ and interpret it as the Euclidean time. We then glue a gravitating region $\CM_G$ along $\partial \left(\Sigma \times I\right)$ such that $\partial \CM_G = \partial \left(\Sigma \times I\right)$. For the gravity sector, we impose the Dirichlet boundary condition (DBC) at the interface between $\CM_G$ and $\Sigma \times I$, while for the QFT sector, the interface is transparent and it lives across the whole spacetime. Let us use $\CM = \CM_G \cup (\Sigma \times I)$ to denote the entire spacetime manifold and $\phi$ to denote the quantum fields. 

We consider the Euclidean GPI with the geometry $\Sigma \times I$ serving as the input data,
\begin{align}\label{eq:CGM}
    Z_{\rm grav}[\Sigma \times I] = \int_{\partial \CM_G = \partial(\Sigma \times I)} \mathcal{D}g_{\mu\nu} \int_{\CM} \mathcal{D}\phi \exp\left(-I_E\right)\,,
\end{align}
where we sum over the QFT configurations on the entire spacetime $\CM$ and the geometries of $\CM_G$. 
This GPI is considered to be computing the ensemble average \footnote{Note that although the analyses performed in \cite{CGM20} do not require the ensemble average, the authors are open to this possibility, and it may in general be necessary to avoid the factorization puzzle \cite{MM04, SSS19}, so we take this more general choice.} of the trace of a family of unnormalized QFT density matrices $\rho_\Sigma \in \CB(\CH_{\Sigma}^{\rm QFT})$ defined on $\tau = 0$, 
\begin{align}\label{eq:CGM_trace}
    Z_{\rm grav}[\Sigma \times I] = \overline{{\Tr \left(\rho_{\Sigma}\right)}} \,,
\end{align}
which is called a gravitationally prepared state. 
Here the overline denotes the ensemble average.
Such a GPI can be dominated by a ``bra-ket wormhole", in which QFT on $\Sigma$ looks entangled with another closed universe and hence mixed at semiclassical level, but one can demonstrate via the replica trick that $\rho_S$ is pure \cite{CGM20}. As a consequence, one may alternatively write \eqref{eq:CGM_trace} as 
\begin{align}
    Z_{\rm grav}[\Sigma \times I] = \overline{\braket{\Psi(\Sigma\times I/\mathbb{Z}_2)|\Psi(\Sigma\times I/\mathbb{Z}_2)}_\Sigma}\,,
\end{align}
where the pure state $\ket{\Psi(\Sigma\times I/\mathbb{Z}_2)}_\Sigma$ is labeled by half of the original input data $\Sigma \times I$, separated by the spatial slice at $\tau = 0$. 

More generally, the inner product, trace, and other index-contractions associated to the quantum states are accordingly defined by cutting and pasting in the non-gravitational region following the same rules as that in QFT.

Now let us generalize this setup to a broader class of GPI, in which a general spacetime subregion is frozen. In particular, while the frozen subregion $\Sigma \times I$ discussed above includes the full Cauchy slice $\Sigma$, now we would like to consider a general $\CM_F$ which only covers a spatial subregion. More precisely, let $\CM_F$ be a $D$-D manifold and consider the GPI 
\begin{align}\label{eq:PFGPI}
    Z_{\rm grav}[\CM_F] = \int_{\partial \CM_G = \partial\CM_F} \mathcal{D}g_{\mu\nu} \int_{\CM} \mathcal{D}\phi \exp\left(-I_E\right)\,. 
\end{align}
Let $F$ be a $(D-1)$-D spatial slice of $\CM_F$, we assume the partially frozen GPI computes the trace of an unnormalized reduced transition matrix $\mathcal{T}_{F^\circ \cup \partial F}$ on $\mathcal{H}_{F^\circ} \otimes \mathcal{H}_{\partial F}$, averaged,
\begin{align}\label{eq:PFGPI_dictionary}
    Z_{\rm grav}[\CM_F] = \overline{{\Tr \left(\mathcal{T}_{F^\circ \cup \partial F}\right)}} \,,
\end{align}
considering its similarity to the reduced transition matrix in QFT \cite{NTTTW20}. Here,
\begin{align}
    F^\circ \equiv F\backslash \partial F\,,
\end{align}
denotes the interior of $F$. 
We also assume that when $F$ is a time-reflection symmetric slice of $\CM_F$,  $\mathcal{T}_{F^\circ \cup \partial F}$ becomes a density matrix $\rho_{F^\circ \cup \partial F}$. The reason why we carefully separate $F^\circ$ and $\partial F$ will become clear soon later. 

The partially frozen GPI in \eqref{eq:PFGPI} provides a subregion generalization of the gravitationally prepared states defined in \cite{CGM20}. Let us then consider if such a state is pure or mixed via the replica trick.

~\par
{\bf Pure nature of the subregion state.---}
Consider the case where $\CM_F$ is time-reflection symmetric with respect to $F$. 
Prepare two copies of $\CM_F$, and call them $\CM_F^{(1)}$ and $\CM_F^{(2)}$ respectively. Cut each of $\CM_F$ along the spatial slice $F$, and denote the two parts separated by it $\CM_{F}^+$ and $\CM_{F}^-$. Now, with four pieces $\CM_F^{(1)\pm}$ and $\CM_F^{(2)\pm}$, let us combine them in the following two ways. First, gluing $\CM_F^{(1)+}$ ($\CM_F^{(2)+}$) and $\CM_F^{(1)-}$ ($\CM_F^{(2)-}$) back to $\CM_F^{(1)}$ ($\CM_F^{(2)}$), the GPI with $\CM_F^{(1)}\sqcup\CM_F^{(2)}$ frozen computes 
\begin{align}
    Z_{\rm grav}[\CM_F^{(1)}\sqcup\CM_F^{(2)}] = \overline{\left({\Tr \left(\rho_{F^\circ \cup \partial F}\right)}\right)^2} \,. 
\end{align}
On the other hand, gluing $\CM_F^{(1)+}$ ($\CM_F^{(2)+}$) and $\CM_F^{(2)-}$ ($\CM_F^{(1)+}$) along $F$ to obtain $\CM_F^{(1)+(2)-}$ ($\CM_F^{(2)+(1)-}$), the GPI with $\CM_F^{(1)+(2)-} \sqcup \CM_F^{(2)+(1)-}$ frozen computes 
\begin{align}
    Z_{\rm grav}[\CM_F^{(1)+(2)-} \sqcup \CM_F^{(2)+(1)-}] = \overline{{\Tr \left(\mathcal{\rho}_{F^\circ \cup \partial F}^2\right)}} \,. 
\end{align}
Since $\CM_F^{(1)}\sqcup\CM_F^{(2)} = \CM_F^{(1)+(2)-} \sqcup \CM_F^{(2)+(1)-}$, we have 
\begin{align}
    \overline{\left({\Tr \left(\rho_{F^\circ \cup \partial F}\right)}\right)^2} = \overline{{\Tr \left(\rho_{F^\circ \cup \partial F}^2\right)}} \,. 
\end{align}
If there is no ensemble average, this is already sufficient for claiming ${\rm rank}\left(\rho_{F^\circ \cup \partial F}\right) = 1$. In cases there is a nontrivial ensemble average, we need a little bit more effort. 

Similar to the 2-replica case discussed above, by considering $n$-replica, we have 
\begin{align}
    \overline{\left({\Tr \left({\rho}_{F^\circ \cup \partial F}\right)}\right)^n} = \overline{{\Tr \left({\rho}_{F^\circ \cup \partial F}^n\right)}} \,. 
\end{align}
Furthermore, by considering different combinations of $2n$-replica, we have 
\begin{align}
    \overline{\left[{\Tr \left(\rho_{F^\circ \cup \partial F}^n\right)} - \left({\Tr \left(\rho_{F^\circ \cup \partial F}\right)}\right)^n\right]^2} = 0 \,. 
\end{align}
As a result, for each member of the ensemble
\begin{align}
    {\left({\Tr \left(\rho_{F^\circ \cup \partial F}\right)}\right)^n} = {{\Tr \left(\rho_{F^\circ \cup \partial F}^n\right)}} \,, 
\end{align}
and therefore ${\rm rank}\left(\rho_{F^\circ \cup \partial F}\right) = 1$. This implies the partially frozen GPI $Z_{\rm grav}[\CM_F]$ prepares a pure state on the spatial subregion $F$. 

As a consequence, in this case, we can refine \eqref{eq:PFGPI_dictionary} as
\begin{align}
    Z_{\rm grav}[\CM_F] =  \overline{\braket{\Psi(\CM_F/\mathbb{Z}_2)|\Psi(\CM_F/\mathbb{Z}_2)}_{F^{\circ} \cup \partial F}}\,.
\end{align}
where $\CM_F/\mathbb{Z}_2 = \CM_F^+ = \CM_F^-$.

Let us briefly discuss the semiclassical interpretation of this result. The subregion generalization \eqref{eq:PFGPI} differs from \eqref{eq:CGM} in an essential way: while $\rho_{\Sigma}$ is defined on a fully frozen time slice, $\rho_{F^\circ \cup \partial F}$ contains an interface between the frozen region $F$ and gravitating region $G$. 
Had we instead cut the frozen spacetime region $\CM_F$ only along $F^{\circ}$, the path integral would have prepared a mixed QFT state on $F^{\circ}$, as in \cite{DQSY20} and in the spirit of \cite{AHMST19,PSSY19,ILM24}. 
The pure state on $F^\circ \cup \partial F$ can therefore be viewed as a purification of this QFT mixed state by a holographic quantum system living on $\partial F$.

Although we have taken $\CM_F$ to be $D$-D, i.e. codimension-0, and considered $(D-1)$-D spatial slice $F$, higher-codimension cases can be obtained by taking certain limits. For instance, let let $\CM_F$ be the near-boundary region of AdS outside a finite radial cutoff, and then send the cutoff to infinity. This simply imposes DBC at the asymptotic boundary of AdS, and the identification \eqref{eq:PFGPI_dictionary} reduces to the standard AdS/CFT dictionary \cite{GKP98,Witten98}. 
In this case, the pure-state nature statement is just saying that the GPI prepares something regarded as a pure state in the dual CFT. This is not how the original Maldacena's formulation of AdS/CFT \cite{Maldacena97} stated, since there the CFT is not literally placed on the asymptotic boundary, but it is a natural interpretation of the GPI dictionary \cite{GKP98,Witten98}. One may similarly consider freezing codimension-2 (or higher) defects in GPI, as in wedge holography \cite{AKTW20,GKPRRRS22} and more generally discussed in \cite{Wei25}. 

In the following, we consider the case where $\CM_F$ is time-reflection symmetric, and study the corresponding pure state $\ket{\Psi(\CM_F/\mathbb{Z}_2)}_{F^{\circ}\cup \partial F}$. For simplicity, we will use the shortened notation $\ket{\Psi}_{F^{\circ}\cup \partial F}$ in the following.

~\newpage
{\bf Entanglement entropy.---} 
Consider the case in which the partially frozen GPI $Z[\CM_F]$ is dominated by a single saddle geometry $\CM = \CM_G \cup \CM_F$ such that it admits a simple semiclassical description \footnote{This requirement is not satisfied by arbitrary choices of $\CM_F$. In particular, not all the finite-cutoff DBCs are elliptic \cite{Witten18}, and the corresponding boundary-value problem can suffer from failures of local uniqueness as well as linear obstructions to the existence of solutions. We will however not worry about these cases in this Letter.}. 

In this case, $\CM = \CM_G \cup \CM_F$ should also admit a time-reflection symmetry whose fixed locus includes $F$. We denote the time-reflection-symmetric spatial slice by $\Sigma$, with $\Sigma \supseteq F$. The complement of $F$ in $\Sigma$ is denoted by $G\equiv\Sigma\backslash F$, where $G$ stands for ``gravitating''. In what follows, we use the convention 
\begin{align}
    F = F^\circ \cup \partial F\,, ~~~ G=G^{\circ} = \Sigma \backslash F\,,
\end{align}
so that $F$ always includes its boundary $\partial F$ whereas $G$ does not. This convention also emphasizes that the interface between $F$ and $G$ is frozen. 

Under the semiclassical approximation, the spatial slice $\Sigma$ as well as the QFT state $\ket{\psi}_{\Sigma}^{\rm QFT}$ living on it together describes a pure state $\ket{\Psi}_F \in \CH_F =\CH^{\rm QFT}_{F^{\circ}} \otimes \CH^{\rm hol}_{\partial F}$ on $F$, analogous to that in the AdS/CFT where a bulk geometry together with the QFT state on it describes a boundary CFT state. 
For a bipartition $F= A\cup B$ with Hilbert space factorization $\mathcal{H}_F = \mathcal{H}_A \otimes \mathcal{H}_B$, we can define the entanglement entropy
\begin{align}
    S_A^{|F} = -\Tr\left(\tilde{\rho}_A^{|F} \log\left(\tilde{\rho}_A^{|F}\right)\right)\,, 
\end{align}
where $\tilde{\rho}_A^{|F} = {\rho}_A^{|F}/{\rm Tr}({\rho}_A^{|F})$ is the normalized counterpart of 
$\rho_A^{|F} \equiv {\Tr}_{B} |\Psi\rangle_F\langle \Psi|_F$. 
Here, we explicit write down the frozen region $F$ in the superscript, in order to emphasize that $\rho_A^{|F}$ is a state relative to the frozen $F$, and also leave open the possibility that the same combination of the geometry $\Sigma$ and bulk QFT state $\ket{\psi}_{\Sigma}^{\rm QFT}$ may realize a different pure state $\ket{\Psi'}_{F'}$ on a different frozen subregion $F'$, which reduces to a different density matrix $\rho_A^{|F'}$. This will be discussed in an upcoming paper \cite{Wei26}.

\begin{figure}
    \centering
    \includegraphics[width=7cm]{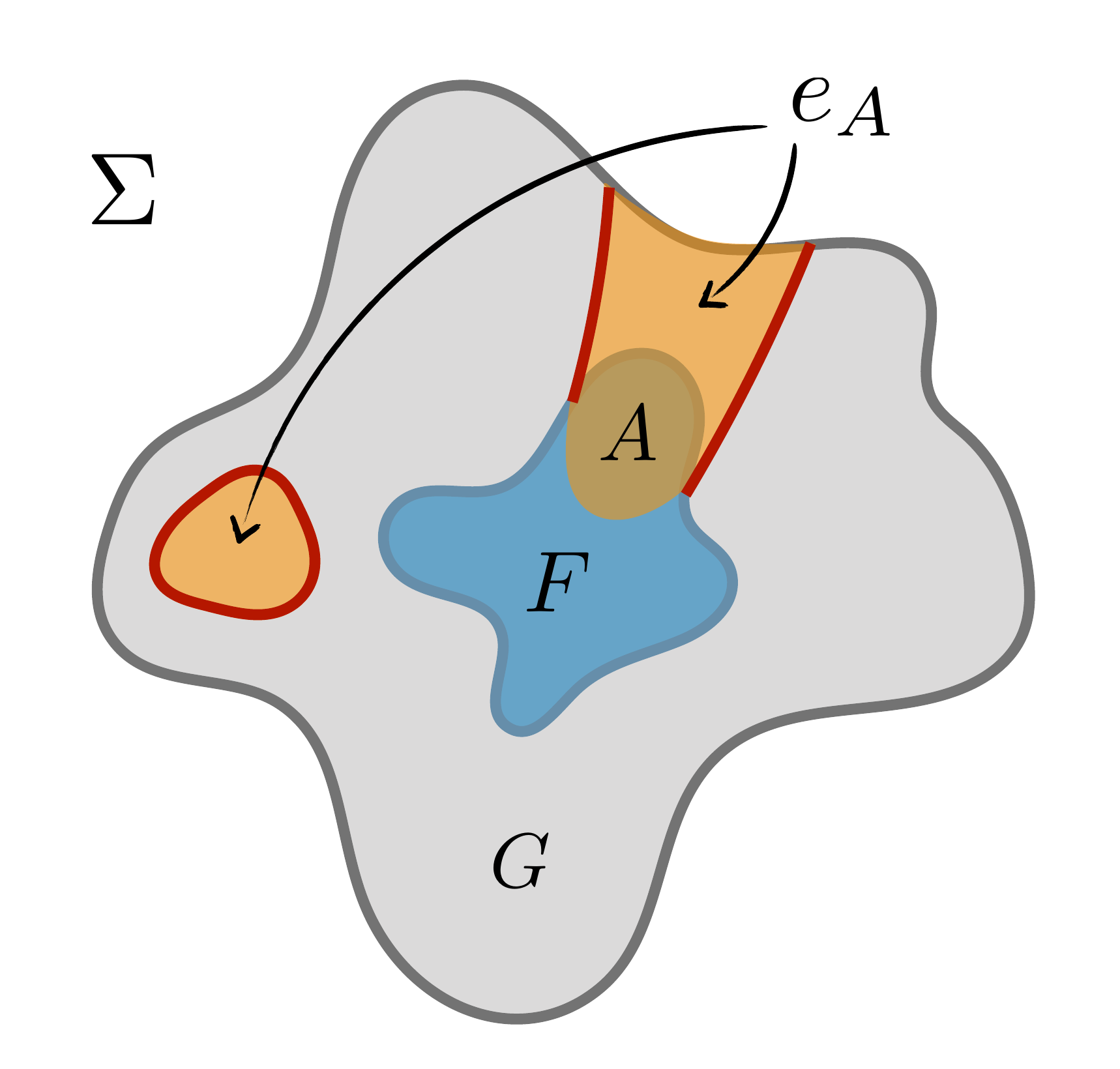}
    \caption{Sketch of the computation of the entanglement entropy using \eqref{eq:ODEE}. Shown is a spatial slice $\Sigma$ with nontrivial gravitating boundary $\partial\Sigma$.  The frozen subregion $F$ is shown in blue, and the ambient  gravitating region $G$ is shown in gray. Given a subregion $A\subset F$, a representative $e_A$ satisfying the frozen condition \eqref{eq:homology} is shown in yellow. The red surface $\gamma^{|F}(e_A)$ is the codimension-2 surface defined in \eqref{eq:gamma}.}
    \label{fig:EW}
\end{figure}

We now seek a holographic prescription to compute $S_A^{|F}$. 
When $\Sigma$ is asymptotically AdS and $F$ is chosen as the region beyond a bulk IR cutoff, it should reduce to the Ryu-Takayanagi (RT) formula \cite{RT06,RT06b,LM13,FLM13,EW15}. 
Considering that in the GPI derivation of the RT formula \cite{LM13}, the candidate RT surface can explore the gravitating region $G$ with constraints induced from the frozen region $F$, it is natural to conjecture the following holographic EE formula: 
\begin{align}\label{eq:ODEE}
    S_A^{|F} = \min_{\substack{e_A \cap F = A}} \left(\frac{{\rm Area}\left(\gamma^{|F}(e_A)\right)}{4G_N} + S_{\rm QFT} (e_A) \right)\,, 
\end{align}
where the minimization is over all regions $e_A$ such that $A \subseteq e_A \subseteq \Sigma$, satisfying
\begin{align}\label{eq:homology}
    e_A \cap F = A\,,
\end{align}
$\gamma^{|F}(e_A)$ is the codimension-2 (i.e. $(D-2)$-D) surface defined as 
\begin{align}\label{eq:gamma}
    \gamma^{|F}(e_A) \equiv {\rm cl}({e_A})\cap {\rm cl}{(G \backslash e_A)}\,, 
\end{align}
where ${\rm cl}({X}) \equiv X \cup \partial X$ is the closure of $X$. This means only the interface between $e_A$ and other gravitating regions contribute to the area term. $S_{\rm QFT} (e_A)$ is the von Neumann entropy of (the normalized counterpart of) $\rho_{e_A}^{|\Sigma} = {\rm Tr}_{\Sigma \backslash e_A} |\psi\rangle_{\Sigma}\langle \psi|_{\Sigma} $, the density matrix of the QFT living on $e_A$.

The frozen nature of $F$ is reflected by the fact that $e_A$ can explore $G$ while its overlap with $F$ is fixed. Hence we will call \eqref{eq:homology} the frozen condition relative to $F$, which is a generalization of the homology condition \cite{HT07} in the RT formula. See Fig. \ref{fig:EW} for a sketch. 

When the $e_A$ minimizing \eqref{eq:ODEE} is unique, let us use $E_A^{|F}$ to denote it and $\Gamma_A^{|F}$ to denote the associated $\gamma^{|F}(e_A)$. The $E_A^{|F}$ and $\Gamma_A^{|F}$ are then a generalization of the entanglement wedge \cite{CKNvR12} and the quantum RT surface \cite{EW15}. An important insight is that the entanglement wedge of $A$ depends on the choice of the frozen subregion $F$. Especially, for the same subregion $A$, whether a point $p\in \Sigma$ is in $A$'s entanglement wedge or not depends on how $F$ is chosen. More on this will be elaborated in \cite{Wei26}. 

In order for $S_A^{|F}$ and ${E}_A^{|F}$ determined by \eqref{eq:ODEE} to be consistent with the quantum state interpretation associated to $\ket{\Psi}_{F}$, 
they need to satisfy some nontrivial properties. 
Some of them are formulated and proven below. 
\begin{itemize}
    \item (Consistency for pure states) When $A=F$, $E_A^{|F}$ is identical to $\Sigma$ and  \eqref{eq:ODEE} gives $S_A^{|F}=0$. 
    \item (No-cloning) If $A \cap B = \varnothing$, then ${\rm int}({E}_A^{| F}) \cap {\rm int}({E}_B^{| F}) = \varnothing$. 
    \item (Strong subadditivity) For three non-overlapping subregions $A,B,C$ in $F$, $S^{|F}_{ABC}+S^{|F}_{B} \leq S_{AB}^{|F} + S_{BC}^{|F}$.
    \item (Entanglement wedge nesting) If two subregions of $F$ are $A\subset B$, then $E_A^{|F} \subset {E}_B^{|F}$. 
    \item (Complementarity) If $A \cup B = F$, ${\rm int}(A) \cap {\rm int}(B) = \varnothing$, then $E_A^{|F} \cup E_B^{|F} = \Sigma$, $S_A^{|F} = S_B^{|F}$ and ${\rm int}(E_A^{|F}) \cap {\rm int}(E_B^{|F}) = \varnothing$. 
\end{itemize}
The consistency for pure states and the complementarity follow directly from the uniqueness of $E_A^{|F}$ and the pure-state nature of $\ket{\psi}_\Sigma^{\rm QFT}$. Proofs of other statements are shown below.

Let us start with no-cloning. Let us assume that ${\rm int}({E}_A^{| F}) \cap {\rm int}({E}_B^{| F}) \neq \varnothing$ and aim to derive a contradiction from it. Due to the homology condition $e_A \cap F = A$ relative to $F$, ${E}_A^{| F} \cap {E}_B^{| F} \subseteq G$. Therefore, 
\begin{align}
    &{\rm Area} {(\Gamma_A^{|F})} + {\rm Area} {(\Gamma_B^{|F})} \nonumber\\
    \geq 
    &{\rm Area} {\left(\gamma^{|F}({E_A^{|F}\backslash E_B^{|F}})\right)} + {\rm Area} {\left(\gamma^{|F}({E_B^{|F}\backslash E_A^{|F})}\right)}. 
\end{align}
On the other hand, the SSA for the QFT implies 
\begin{align}
    &S_{\rm QFT} {(E_A^{|F})} + S_{\rm QFT} {(E_B^{|F})} \nonumber\\
    \geq 
    &S_{\rm QFT} {({E_A^{|F}\backslash E_B^{|F}})} + S_{\rm QFT} {({E_B^{|F}\backslash E_A^{|F}})}.
\end{align}
Taking the sum of the two, it implies either ${E}_A^{| F}$ or ${E}_B^{| F}$ does not minimize the right hand side of \eqref{eq:ODEE}, which is a contradiction. Therefore, ${\rm int}({E}_A^{| F}) \cap {\rm int}({E}_B^{| F}) = \varnothing$.

Then let us turn to the SSA. 
This can be directly proven by explicitly writing down 
\begin{align}\label{eq:SSA}
    &S_{AB}^{|F} + S_{BC}^{|F} \nonumber \\
    =& \frac{{\rm Area}(\Gamma_{AB}^{|F})}{4G_N} + S_{\rm QFT}(E_{AB}^{|F}) + \frac{{\rm Area}(\Gamma_{BC}^{|F})}{4G_N} + S_{\rm QFT}(E_{BC}^{|F}) \nonumber \\ 
    \geq& \frac{{\rm Area}(\Gamma_{AB}^{|F})}{4G_N} + \frac{{\rm Area}(\Gamma_{BC}^{|F})}{4G_N}  \nonumber \\
    &+S_{\rm QFT}(E_{AB}^{|F} \cup E_{BC}^{|F}) + S_{\rm QFT}(E_{AB}^{|F} \cap E_{BC}^{|F}) \nonumber \\ 
    \geq& \frac{{\rm Area}\left(\gamma^{|F}(E_{AB}^{|F} \cup E_{BC}^{|F})\right)}{4G_N}  + S_{\rm QFT}(E_{AB}^{|F} \cup E_{BC}^{|F}) \nonumber\\
    &+ \frac{{\rm Area}\left(\gamma^{|F}(E_{AB}^{|F} \cap E_{BC}^{|F})\right)}{4G_N}  + S_{\rm QFT}(E_{AB}^{|F} \cap E_{BC}^{|F}) \nonumber \\
    \geq& S_{ABC}^{|F} + S_{B}^{|F}, 
\end{align}
where in the third line, we have applied the SSA for the matter field, and in the last line, we have applied the fact that $E_{AB}^{|F} \cup E_{BC}^{|F}$ qualifies as a candidate $e_{ABC}$ and $E_{AB}^{|F} \cap E_{BC}^{|F}$ qualifies as a candidate $e_B$. 

The entanglement wedge nesting works similarly. Let us assume $E_A^{|F} \cap {E}_B^{|F} \neq E_A^{|F}$ and aim to derive a contradiction from this assumption. Under this assumption, with a similar procedure to \eqref{eq:SSA}, we have 
\begin{align}\label{eq:nesting}
    &S_A^{|F} + S_B^{|F} \nonumber\\
    =& \frac{{\rm Area}(\Gamma_{A}^{|F})}{4G_N} + S_{\rm QFT}(E_{A}^{|F}) + \frac{{\rm Area}(\Gamma_{B}^{|F})}{4G_N} + S_{\rm QFT}(E_{B}^{|F}) \nonumber \\ 
    \geq& \frac{{\rm Area}(\Gamma_{A}^{|F})}{4G_N} + \frac{{\rm Area}(\Gamma_{B}^{|F})}{4G_N} \nonumber \\
    &+ S_{\rm QFT}(E_{A}^{|F} \cup E_{B}^{|F}) +  S_{\rm QFT}(E_{A}^{|F} \cap E_B^{|F})  \nonumber \\
    \geq& \frac{{\rm Area}\left(\gamma^{|F}(E_{A}^{|F} \cup E_{B}^{|F})\right)}{4G_N}  + S_{\rm QFT}(E_{A}^{|F} \cup E_{B}^{|F}) \nonumber\\
    &+ \frac{{\rm Area}\left(\gamma^{|F}(E_{A}^{|F} \cap E_{B}^{|F})\right)}{4G_N}  + S_{\rm QFT}(E_{A}^{|F} \cap E_{B}^{|F}) 
\end{align}
Since $E_A^{|F} \cup E_B^{|F}$ and $E_A^{|F} \cap E_B^{|F}$ qualify as candidate $e_{B}$ and candidate $e_A$ respectively, \eqref{eq:nesting} implies that either $E_A^{|F}$ or $E_B^{|F}$ does not minimize the right hand side \eqref{eq:ODEE}, which is a contradiction. Therefore, $E_A^{|F} \subseteq {E}_B^{|F}$. Furthermore, due to \eqref{eq:homology}, the frozen condition relative to $F$, $E_A^{|F} \subset {E}_B^{|F}$.

~\par
{\bf Special cases.---}
Besides the reduction to the RT formula in the case described above, let us discuss some other important special cases of \eqref{eq:ODEE}. Note again that the formula \ref{eq:ODEE} is designed for codimension-1 subregions. Therefore, when we say some subregion is codimension-2/codimension-3, we mean it is a codimension-1 subregion thinned to a higher codimension defect in an appropriate regulated limit. 

Let $\Sigma$ be a Cauchy slice of an AdS gravity coupled to a flat nongravitational heat bath at the asymptotic boundary, and $F$ be the heat bath region. For the EE between $F^{\circ}$ and $\partial F$ (specified with a cutoff), \eqref{eq:ODEE} reduces to the island formula \cite{Penington19,AEMM19,AMMZ19,PSSY19,AHMST19}. 
If one instead considers a general subregion $F$ on a general Cauchy slice $\Sigma$ and let $A = F^{\circ}$, then \eqref{eq:ODEE} computes the effective entropy introduced in \cite{DQSY20}, which is a close relative of the island formula.

If $\Sigma$ has nontrivial boundary $\partial \Sigma$, and $F$ is a proper subregion of $\partial \Sigma$ with a cutoff, \eqref{eq:ODEE} reduces to the holographic EE formula in the presence of an end-of-the-world (EOW) brane, identified with $G\cap \partial \Sigma$. When $F$ is taken the limit to be codimension-2, then it matches the RT formula in AdS/BCFT \cite{Takayanagi11,FTT11,HM13}. When $F$ is taken the limit to be codimension-3, it matches the HEE formula in wedge holography \cite{AKTW20,GKPRRRS22}. 

Finally, suppose $\Sigma$ is AAdS and $A$ is a bulk subregion disconnected from $\partial \Sigma$. Choosing $F=\partial\Sigma\cup A$, the region $E_A^{|F}$ coincides with the Bousso--Penington (BP) generalized entanglement wedge of $A$ \cite{BP22,BP23,BC23,KRR25}. 
We note that this realization is however not unique: freezing instead $F'=\partial\Sigma\cup\partial A$, $E_{\partial A}^{|F'}$, supplemented by a choice of side of $\partial A$ corresponding to $A$, gives the same wedge. However, these might not be the point of view taken by the original authors \cite{BP22, BP23}, since they intend to consider fully gravitational subregion, while we need to freeze at least $\partial A$. 

~\par
{\bf Discussion.---} 
In this Letter, we started by generalizing gravitationally prepared states to spatial subregions. 
This construction is a partially frozen GPI, where a codimension-0 spacetime subregion is fixed, with ordinary QFT rules imposed there, and higher-codimension regions arise as limits. We showed that this GPI prepares a pure state on the spatial subregion.
This differs from ordinary quantum many-body systems, where subregions are described by mixed states due to their openness. This pure state is interpreted as a mixed state associated to its interior purified by holographic degrees of freedom localized on its boundary, $\ket{\Psi}_{F^\circ \cup \partial F} \in \mathcal{H}^{\rm QFT}_{F^{\circ}} \otimes \mathcal{H}^{\rm QFT}_{\partial F}$. As a crucial characterization, EE associated with different bipartitions of $F$ is considered. When the partially frozen GPI is dominated by a single geometry, we propose the RT-like prescription \eqref{eq:ODEE} to compute the EE holographically. We checked its consistency for pure states, no-cloning, SSA, entanglement wedge nesting, and complementarity, and explained that it reproduces several known entropy formulas in special cases. 

Along the way of the exploration, an interesting feature becomes clear. 
Such a formulation defines infinitely many nongravitational quantum theories from GPI, all with the same gravitational action but different choices of $F$ and $\CM_F$. In particular, two different frozen spacetime subregions $\CM_F$ and $\CM_{F'}$ associated with two different spatial subregions $F$ and $F'$ can be dominated by the same semiclassical spacetime $\CM$ with the same time-reflection symmetric slice $\Sigma$ and QFT state $\ket{\psi}_\Sigma$. Thus a single semiclassical spacetime can be encoded in different quantum states $\ket{\Psi}_F$ and $\ket{\Psi'}_{F'}$ living in different subregions of $\Sigma$. These may be interpreted as descriptions available to different observers localized at $F$ and $F'$, respectively. Each observer trades their knowledge on local geometric data for a quantum state describing the entire spacetime. In this sense, $S_A^{|F}$ and $E_A^{|F}$ determined via \eqref{eq:ODEE} give observer-dependent generalizations of EE and the entanglement wedge.

The remaining challenge is to isolate the observer-independent content shared among these descriptions, and to formulate precise maps between different observers' viewpoints. We will address this in \cite{Wei26}, where the observer-dependent entanglement wedge introduced here plays a central role.

~\par
{\bf Note added.---} 
After this work was completed and during preparation for submission, an interesting paper \cite{BKLS26} appeared on arXiv, discussing a closely related setup.
In particular, we expect, for pure gravity without matter QFT, taking 
$\CM_F = F \times \text{(infinitesimal interval)}$ in our setup relates it to the 
``Dirichlet data'' setup of \cite{BKLS26}.

\section*{Acknowledgements}
I am grateful to Rapha\"el Dulac, Elliott Gesteau, Aidan Herderschee, Masamichi Miyaji, Andrew Strominger, Diandian Wang, Mengyang Zhang, and especially Daniel Jafferis and Tadashi Takayanagi for helpful discussions and comments. I am supported by the Society of Fellows at Harvard University.

\bibliography{subregion_holography}

\end{document}